\documentclass[useAMS,usenatbib]{mn2e}
\usepackage{times}
\usepackage{graphics,epsfig}
\usepackage{graphicx}
\usepackage{amssymb}
\usepackage{txfonts}

\title[{\rm H}~{\sc i} spin temperature in two $z>3$ DLAs]{Stringent constraints on the H~{\sc i} spin temperature in two $z>3$ Damped Lyman-$\alpha$ systems from redshifted 21~cm absorption studies}
\author[Roy et al.]{Nirupam Roy$^{1}$\thanks{E-mail: nirupam@mpifr-bonn.mpg.de}, Smita Mathur$^{2,3}$, Vishal Gajjar$^{4}$ and Narendra Nath Patra$^{4}$\\ 
$^{1}$Max-Planck-Institut f\"{u}r Radioastronomie, Auf dem H\"{u}gel 69, D-53121 Bonn, Germany\\
$^{2}$Department of Astronomy, The Ohio State University, Columbus, OH 43210, USA\\ 
$^{3}$Center for Cosmology and Astro-Particle Physics, The Ohio State University, Columbus, OH 43210, USA\\ 
$^{4}$National Centre for Radio Astrophysics, Tata Institute of Fundamental Research, Post Bag 3, Ganeshkhind PO, Pune 411007, Maharashtra, India}
\begin{document}

\date{Accepted 2013 August 16. Received 2013 August 16; in original form 2013 April 15}

\pagerange{\pageref{firstpage}--\pageref{lastpage}} \pubyear{2013}

\maketitle

\label{firstpage}

\begin{abstract}
Physical properties of Damped Lyman-$\alpha$ absorbers and their evolution are 
closely related to galaxy formation and evolution theories, and have important 
cosmological implications. H~{\sc i} 21~cm absorption study is one useful 
way of measuring the temperature of these systems. In this work, very strong 
constraints on the temperature of two Damped Lyman-$\alpha$ absorbers at $z > 
3$ are derived from low radio frequency observations. The H~{\sc i} spin 
temperature is found to be greater than $2000$ K for both the absorbers. The 
high spin temperature of these high-redshift systems is in agreement with the 
trend found in a compilation of temperatures for other Damped Lyman-$\alpha$ 
absorbers. We also argue that the temperature -- metallicity relation, 
reported earlier in the literature, is unlikely to be a spurious line of sight 
effect, and that the redshift evolution of the spin temperature does not 
arises due to a selection effect. All of these are consistent with a redshift 
evolution of the warm gas fraction in Damped Lyman-$\alpha$ systems. 
\end{abstract}


\section{Introduction}
\label{sec:intro}

Damped Lyman-$\alpha$ (DLA) absorbers are very important probes of neutral 
hydrogen (H~{\sc i}) in the distant Universe. These high H~{\sc i} column 
density structures (${\rm N}_{\rm HI}\gtrsim2\times10^{20}$ cm$^{-2}$), seen 
against the quasar continuum emission, are believed to contain most of the 
neutral gas at high redshift, and are supposed to be the main sites of star 
formation and the precursors of galaxies. A detailed study of the physical 
conditions of these high column density systems is, therefore, necessary for 
understanding galaxy formation and evolution and the large scale structure of 
the Universe. DLAs are also useful tools to address cosmological problems 
\citep[e.g., the evolution of fundamental constants;][]{wol76,sri10}. 

In spite of the significant amount of work already done on this subject, 
properties of DLAs, such as density, H~{\sc i} temperature, structure and 
size, are still unsettled \citep[e.g.,][]{wol05}. The issue of redshift 
evolution of these properties is also controversial \citep{kan04,wol05,kan07}. 
One particular interesting issue in this regard is the temperature of the 
absorbing gas. Radio frequency observations of redshifted H~{\sc i} 21 cm {\it 
absorption} lines from DLAs, against suitable background radio continuum 
sources (e.g. radio loud quasars), are widely used to constrain the spin 
temperatures ($T_{\rm s}$). It has been argued that the derived values show a 
trend of high $T_{\rm s}$ at higher $z$, but both high and low value of 
$T_{\rm s}$ are observed at low $z$. Processes like collision and resonant 
scattering of Lyman-$\alpha$ photons generally couple $T_{\rm s}$ with the 
kinetic temperature, $T_{\rm k}$ \citep{fie58}. So a higher $T_{\rm s}$ 
implies a higher average $T_{\rm k}$ of the neutral gas, or equivalently a 
smaller fraction of cold gas, at high $z$. If true, this may be a very 
important constraint in galaxy evolution models. However, this trend is so far 
deduced mostly from low redshift measurements and a few lower limits of 
$T_{\rm s}$ based on non-detection of H~{\sc i} 21 cm absorption at higher 
redshift \citep[e.g.,][]{kan03,kan07}. It is also possible that the compact 
reservoirs of H~{\sc i} observed against large radio continuum sources have 
covering factors smaller than the assumed value of unity, so $T_{\rm s}$ 
smaller than those reported \citep{cur05,cur06}. Alternatively, the optical 
and radio lines of sight could be drastically different \citep{wol03}. 
Moreover, there are contradicting evidences of both cold and warm gas in high 
redshift DLAs 
\citep[e.g.,][]{car96,deb96,bri97,wol03,how05,kan07,yor07,jor09,sri10,car12,kan13},
which are not consistent with any strong trend. Given the importance of these 
results in galaxy evolution models, it is necessary to address these issues 
critically and in much greater details.

Naturally, the main problem here is the lack of H~{\sc i} 21~cm absorption 
studies and detections at higher redshift. Most such studies are focused on 
low redshift \citep[e.g.,][]{che00,kan01}. To our knowledge, there are only 
four reported detections of 21~cm absorption at $z = 2 - 3$ 
\citep{wol85,kan06,yor07,kan09b,kan13} and just two cases of confirmed 
detection at $z>3$: towards PKS~$0201+113$ 
\citep[$z=3.388$,][]{deb96,bri97,kan07} and towards J$1337+3152$ 
\citep[$z=3.174$,][]{sri10,sri12}. In light of this, we have started a program 
of very deep observations of high redshift DLAs with the Giant Meterwave Radio 
Telescope \citep[GMRT;][]{swa91} in an attempt to increase the number of such 
observations at $z>3$; our goal is to constrain $T_{\rm s}$ and to address 
some of the above mentioned issues systematically. High sensitivity of the 
telescope, and the spatial and spectral resolution of the instrument make GMRT 
suitable for this project. Note that the GMRT 325 MHz receiver covers the 
H~{\sc i} 21~cm line for a redshift range of $z \approx 3.2 - 3.6$. The Sloan 
Digital Sky Survey (SDSS) damped Lyman-$\alpha$ survey catalogue 
\citep{pro05,pro09} is used to identify a total of $63$ sources in this 
redshift range and $\log({\rm N}_{\rm HI}) \gtrsim 20.7$. Out of these, $7$ 
have radio continuum counterpart detected in the NRAO VLA Sky Survey 
\citep[NVSS;][]{con98}, so are suitable for H~{\sc i} 21~cm absorption 
studies. In this {\it Letter} we report results for two sources from this 
sample for which we have carried out deep GMRT observations and analysis. 
These sources are J$080137.68+472528.2$ and J$001115.23+144601.8$. In 
\S\ref{sec:obs}, the details of the observations and data analysis are 
outlined, while the results are in \S\ref{sec:rslt}. A brief discussion on the 
implications of the results are in \S\ref{sec:dis}, and the conclusions are 
presented in \S\ref{sec:con}.

\section{Targets, Observations and data analysis} 
\label{sec:obs}

J$080137.68+472528.2$ is a quasar at $z_{em} = 3.27558$, with a 1.4 GHz flux 
density of $78.8$ mJy (from NVSS). The DLA system towards this background 
source is at a redshift of $z_{abs} = 3.2228$ with $\log({\rm N}_{\rm HI}) = 
20.70$. The second source, J$001115.23+144601.8$ is at $z_{em} = 4.96717$ with 
NVSS $S_{1.4\ {\rm GHz}} = 35.8$ mJy, and the DLA system is at $z_{abs} = 
3.4523$ with $\log({\rm N}_{\rm HI}) = 21.65$. For the first source, the flux 
density at $325$ MHz from the Westerbork Northern Sky Survey 
\citep[WENSS;][]{ren97} is $\sim 160$ mJy, but the field of the second source 
was not covered by WENSS.

The GMRT observations were carried out during observing cycle 20 (Project ID 
20\_026) on 2011 May 20-27. A total of 72 hours of observing time was 
scheduled for these two sources, but some of the observing runs were affected 
by bad weather. As a result, additional 24 hours of reobservation time was 
scheduled on 2012 February 7-8. The effective on-source time was $\sim 30$ 
hours per source, and the rest (about $15\%$) was spent on observation setup 
and calibration overheads.

For these observations, we used the GMRT 325 MHz receiver with a baseband 
bandwidth of 1 MHz (centred at the redshifted H~{\sc i} frequency for the 
corresponding $z_{abs}$ value) divided into 256 spectral channels. This 
resulted in a spectral resolution of $\sim 3.6$ km s$^{-1}$ and a total 
bandwidth coverage of $\sim 900$ km s$^{-1}$. Standard calibrators, 3C~147, 
3C~286, and 3C~48, were observed to calibrate the flux density scale. Two more 
calibrator sources, $0834+555$ and $2340+135$, were also observed as phase 
calibrators for J$080137.68+472528.2$ and J$001115.23+144601.8$ respectively.

Standard data reduction and analysis were carried out using the Astronomical 
Image Processing System of the National Radio Astronomy Observatory (NRAO 
{\small AIPS}). Bad data were carefully flagged before flux, phase and 
bandpass calibrations. A very stable spectral bandshape is required to detect 
any weak H~{\sc i} absorption line. Hence, the strong phase calibrators, along 
with all the flux calibrators, were also used for bandpass calibration, and an 
interpolated bandpass solution was applied to the target source data. The 
flagged and calibrated data for the same source from multiple days of 
observing runs were then combined together and self-calibrated using the 
initial continuum image. The final continuum flux density was found to be $ 
\sim 169$ mJy, in good agreement with the WENSS flux density, for 
J$080137.68+472528.2$.  For J$001115.23+144601.8$, however, we found two more 
point sources very close to the target (within the NVSS synthesized beam, and 
hence unresolved by NVSS); the flux density of the target quasar is only $\sim 
24$ mJy. The continuum subtracted data were then used to produce the spectral 
image cubes, and the spectra were extracted for both the lines of sight from 
the corresponding positions of the unresolved continuum sources. Finally, the 
measured continuum flux densities were used to convert the flux density 
spectra to optical depth spectra assuming a covering factor of unity.

\section{Results}
\label{sec:rslt}

\begin{figure*}
\begin{center}
\includegraphics[scale=0.33,angle=-90.0]{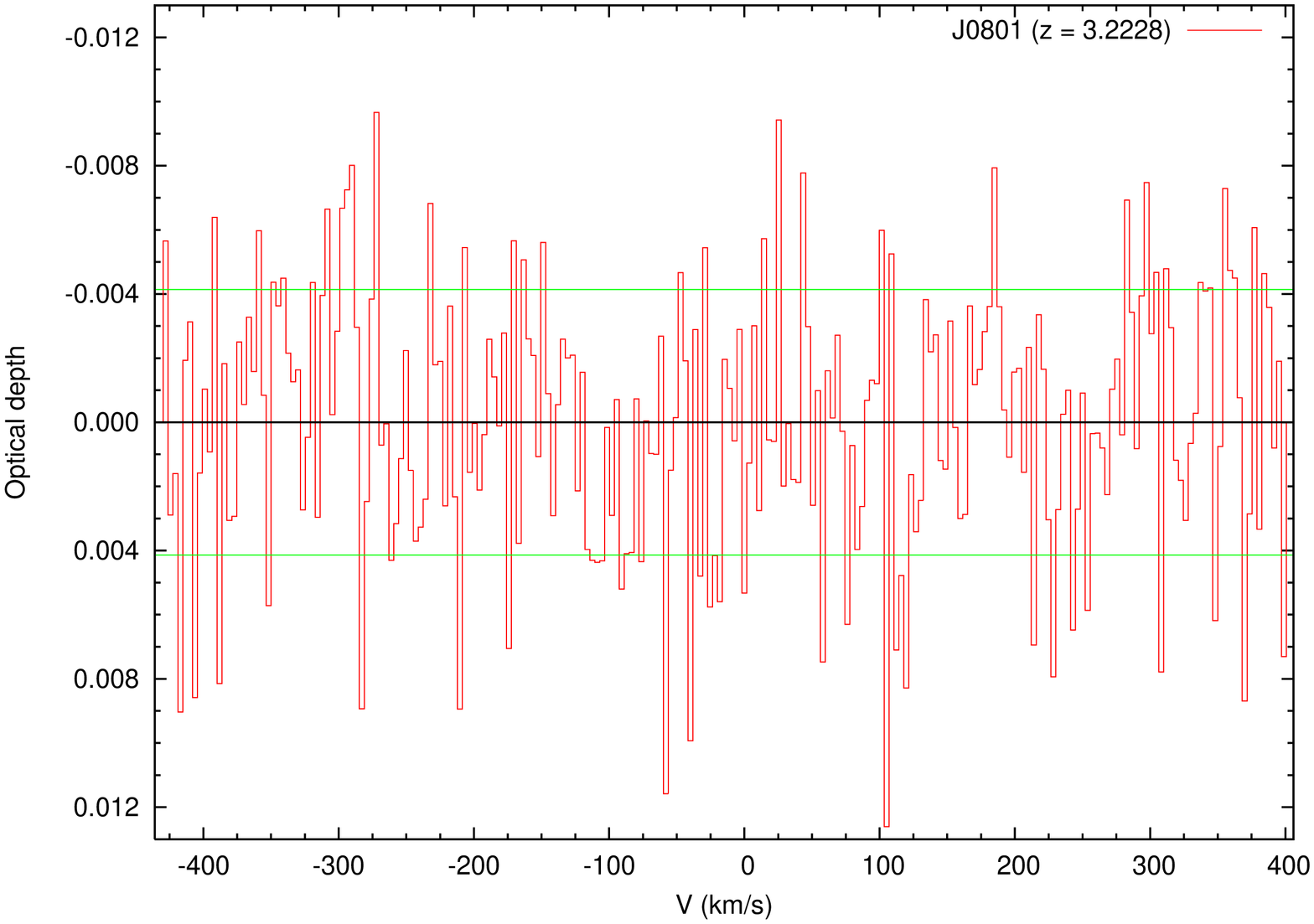}
\includegraphics[scale=0.33,angle=-90.0]{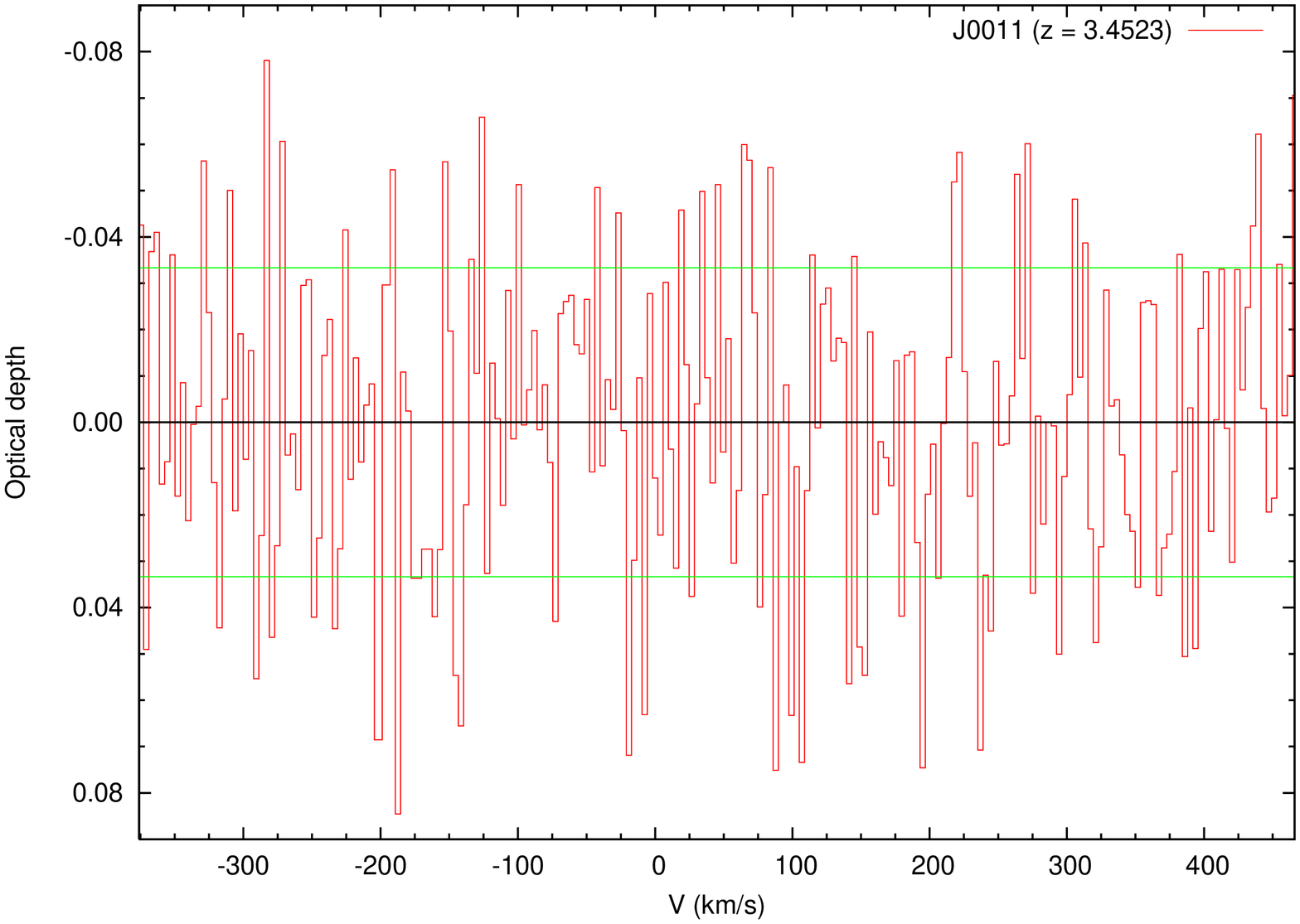}
\caption{\label{fig:specs} The observed optical depth spectra towards J$080137.68+472528.2$ (left) and J$001115.23+144601.8$ (right) showing non-detection of H~{\sc i} 21~cm absorption from DLAs at $z = 3.2228$ and $3.4523$ respectively. The (green) horizontal lines show the $1\sigma$ range of optical depths $\pm 0.004$ and $\pm 0.033$ for the two spectra respectively.}
\end{center}
\end{figure*}

The final spectra for both the lines of sight are shown in 
Figure~\ref{fig:specs}. With the long integration time on each source, we 
reached very good (sub mJy per $< 4$ km~s$^{-1}$) RMS noise sensitivity 
limits, but H~{\sc i} 21~cm absorption is not detected towards either of these 
lines of sight. The spectral RMS noise values are $\sim 0.83$ mJy per $\sim 
3.8$ km~s$^{-1}$ for J$001115.23+144601.8$, $\sim 1.2$ mJy (May 2011) and 
$0.88$ mJy (Feb. 2012) per $\sim 3.6$ km~s$^{-1}$ for J$080137.68+472528.2$. 
The final RMS optical depths at these velocity resolutions are $0.004$ and 
$0.033$ for J$080137.68+472528.2$ and J$001115.23+144601.8$ respectively. 
Based on the known H~{\sc i} column densities from SDSS for these DLA systems, 
one can now constrain the spin temperature $T_{\rm s}$. We use the standard 
relation
\begin{equation}
{\rm N}_{\rm HI} = 1.823\times10^{18} \frac{{T}_{\rm s}}{f_c}\int\tau{dv}
\label{eqn:eqn1}
\end{equation}
where $f_c$ is the covering factor, and assume that the integral is over a 
Gaussian profile with a full width at half maximum (FWHM) of $\Delta v$ 
km~s$^{-1}$. The RMS optical depths finally translate to $3\sigma$ limits of 
$T_{\rm s}/f_c > 3453 \times (10/\Delta v)^{1/2}$ and $> 3723 \times 
(10/\Delta v)^{1/2}$ K for J$080137.68+472528.2$ and J$001115.23+144601.8$ 
respectively.

Neither of the target sources, unfortunately, have reported Very Long Baseline 
Interferometry (VLBI) observations, which are critical to measure $f_{c}$ and 
break the degeneracy between $T_{\rm s}$ and $f_{c}$. However, based on a 
sample of $26$ background radio sources with foreground DLAs, 
\citet{kan09a,kan13} reported a median covering factor of $0.625$, with a 
minimum value of $0.3$ and no significant trend with $z$. Adopting $f_c = 
0.625$ and $\Delta v = 10$ km~s$^{-1}$, the current non-detection limits will 
imply $T_{\rm s} > 2158$ and $> 2326$ K for these two systems. Even for $f_c 
= 0.3$, $T_{\rm s}$ for both the systems are above $1000$ K. Note that an 
FWHM $\approx 10$ km~s$^{-1}$ corresponds to pure thermal broadening for 
$\sim 2000$ K gas.

\begin{figure}
\begin{center}
\includegraphics[scale=0.33,angle=-90.0]{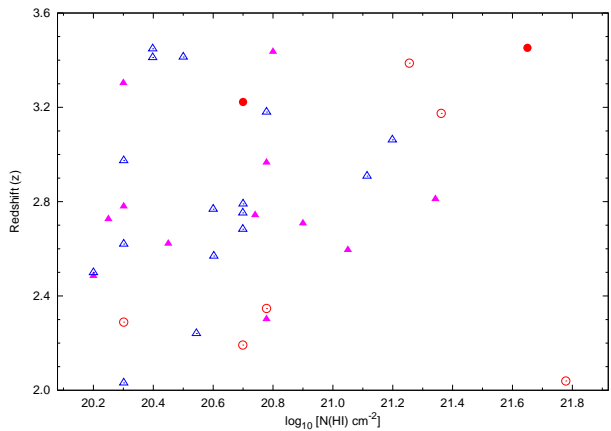}
\caption{\label{fig:fig2} All $z>2$ DLAs with $T_{\rm s}$ measurements or limits from the literature in redshift -- column density parameter space. The open circles are the only six cases of detection of H~{\sc i} absorption (with known $f_c$). Triangles are sources with $T_{\rm s}$ limits: open triangles for known $f_c$ and filled triangles for unknowns. Filled circles are the two from this work (with unknown $f_c$).}
\end{center}
\end{figure}

To compare the present limits with measurements or limits of $T_{\rm s}$ from 
earlier similar observations, we have compiled information from the literature 
reporting the redshifted H~{\sc i} 21~cm absorption studies of high redshift 
($z > 2$) DLAs. In the redshift range of $2.031 - 3.4523$ (that is, the age of 
the universe $\sim 1.9 - 3.3$ Gyr for standard cosmological parameters H$_0 = 
71$ km~s$^{-1}$Mpc$^{-1}$, $\Omega_{\rm M} = 0.27$ and $\Omega_\Lambda = 
0.73$), we found $6$ reported detections and $30$ (including the two from this 
work) non-detections of H~{\sc i} absorption 
\citep[][]{wol85,car96,deb96,bri97,kan03,kan06,kan07,yor07,kan09b,cur10,sri10,sri12,kan13} 
with the value (or the limit) of $T_{\rm s}$ (or $T_{\rm s}/$$f_{c}$) in the 
literature. For all six cases of detection and $16$ of the non-detections, the 
covering factor $f_c$ is also available. The median value of $f_c$ is $0.65$ 
for this sample of $22$ sources, and there is, again, no statistical trend 
with redshift. For the other $14$ sources, we used the median value of $f_c = 
0.65$ to estimate, in a statistical sense, the expected $T_{\rm s}$ limits 
from the $T_{\rm s}/$$f_{c}$ limits. If, for a given source, multiple 
estimates are available in the literature, we used the most updated, sensitive 
and best estimate. Figure~\ref{fig:fig2} shows the redshift and column density 
distribution of the whole sample. We note a selection trend here, though based 
on a small number statistics: radio observations for H~{\sc i} absorption 
studies appear to have targeted low N(H~{\sc i}) systems at the lower end of 
the $z > 2$ redshift range, but both low and high N(H~{\sc i}) systems at 
higher redshifts. It is worth remembering that the target selection criteria 
are very different in different studies, and hence the collated sample is 
expected to be significantly heterogeneous as well as incomplete.

\begin{figure*}
\begin{center}
\includegraphics[scale=0.33,angle=-90.0]{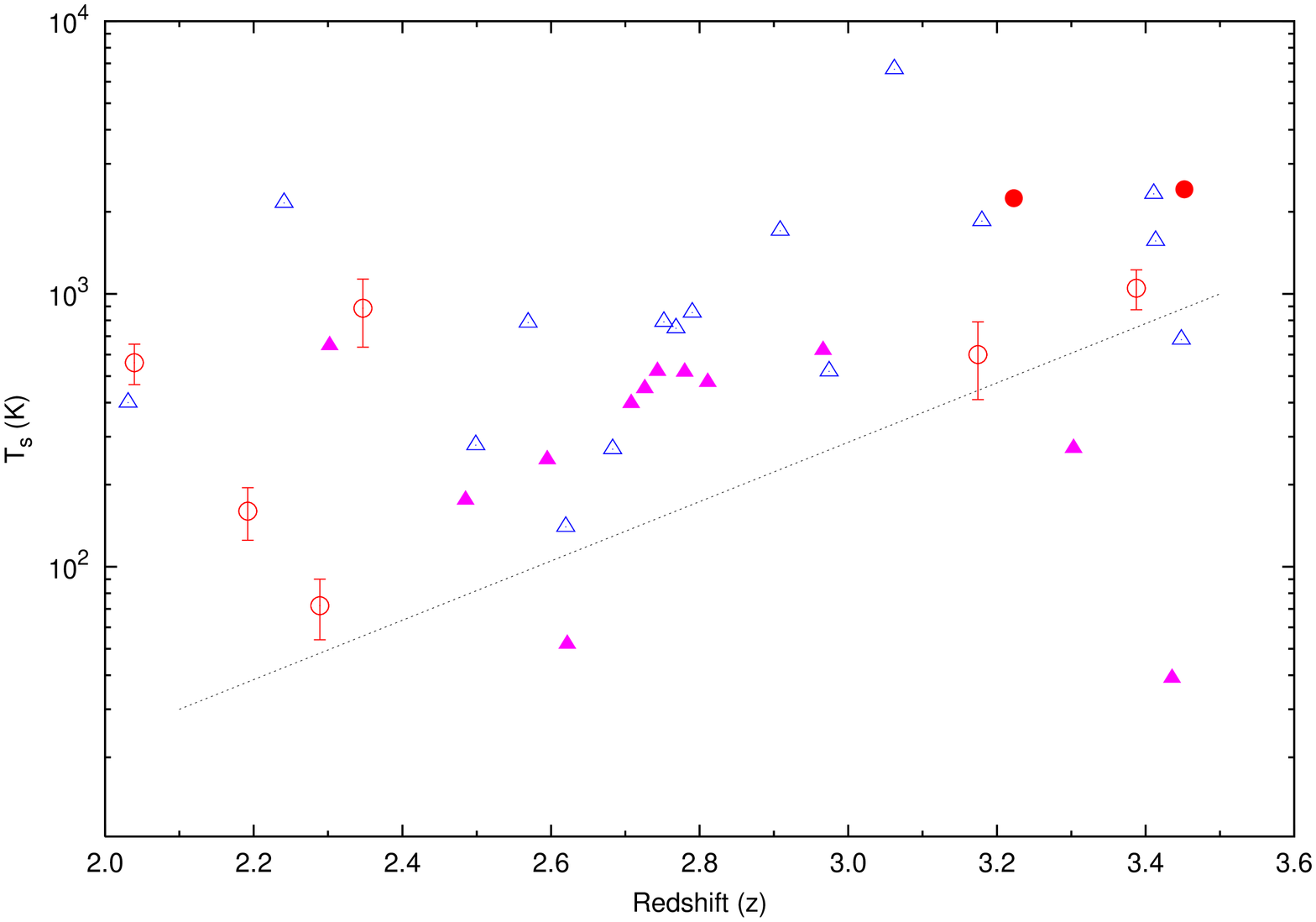}
\includegraphics[scale=0.33,angle=-90.0]{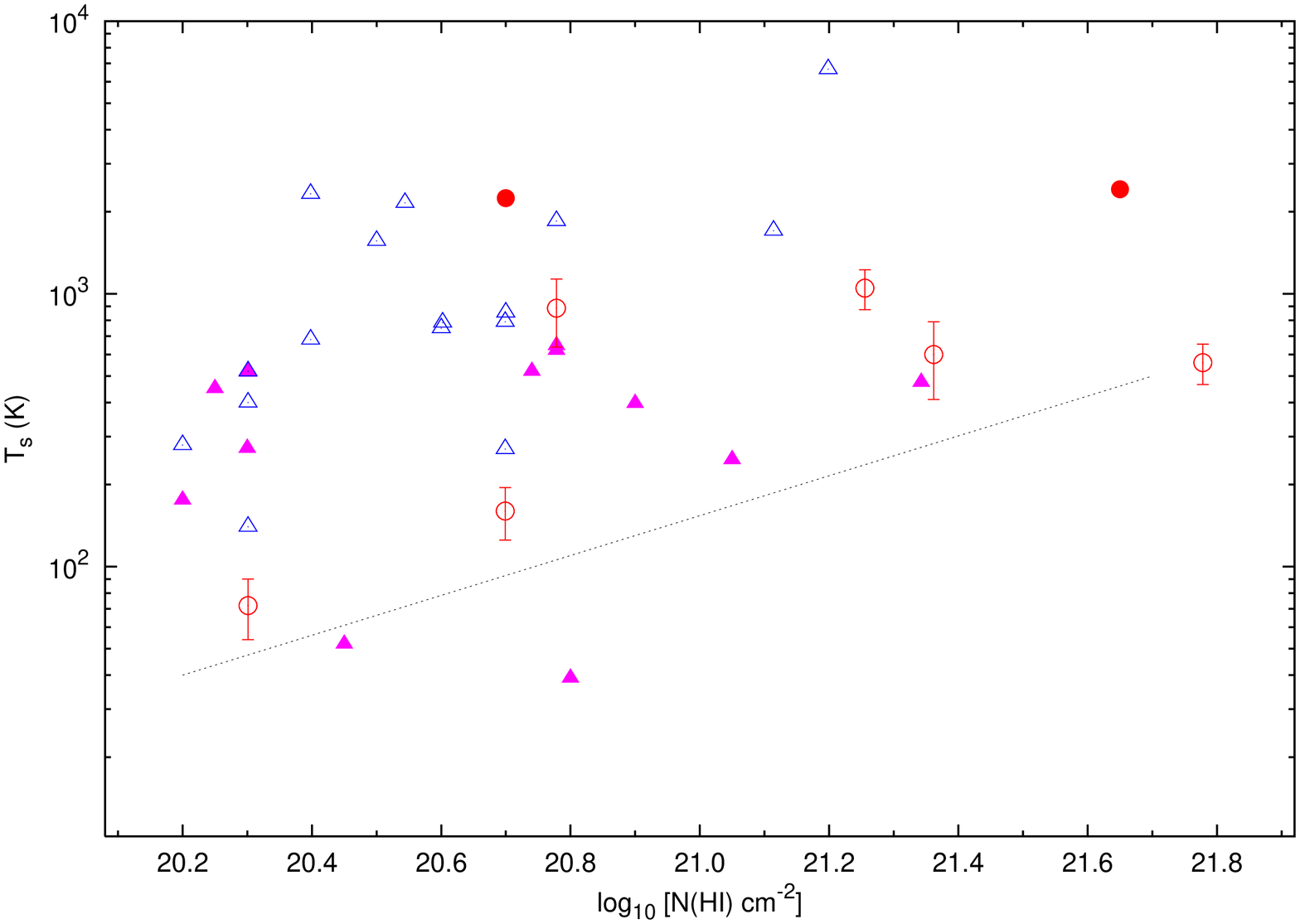}
\caption{\label{fig:fig3} $T_{\rm s}$ measurements or lower limits as a function of redshift and H~{\sc i} column density for the $z>2$ DLA sample. Legends are same as in Figure~\ref{fig:fig2}. The detections are shown with standard errors for $T_{\rm s}$. For unknown $f_c$, the median value ($f_c = 0.65$) of the known $f_c$ sub-sample is adopted to statistically estimate the expected $T_{\rm s}$ limit from the $T_{\rm s}/$$f_{c}$ limit. The dotted lines indicate plausible minimum $T_{\rm s}$ at different $z$ and N(H~{\sc i}). See \S\ref{sec:rslt} and \S\ref{sec:dis} for details.}
\end{center}
\end{figure*}

\begin{figure*}
\begin{center}
\includegraphics[scale=0.33,angle=-90.0]{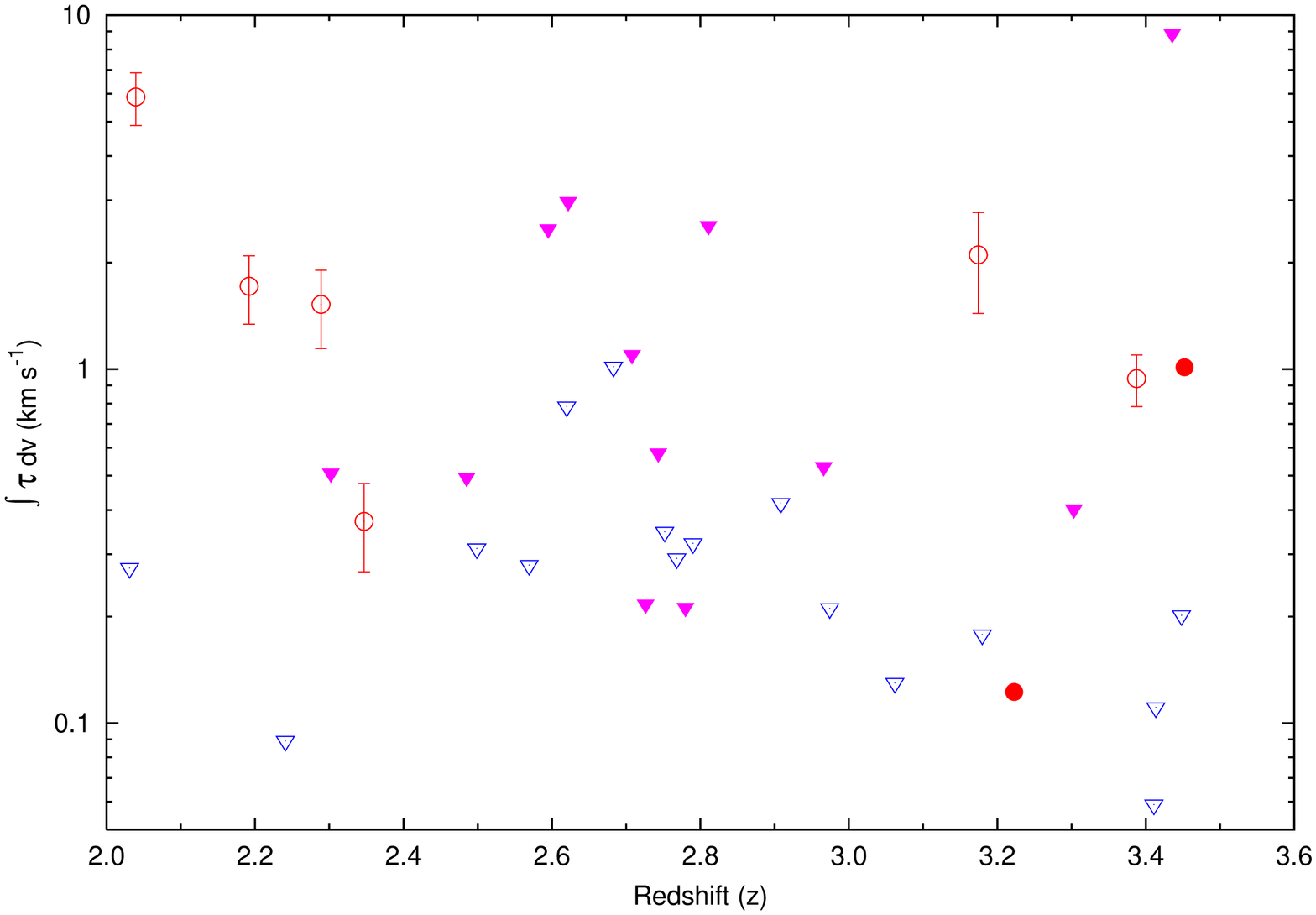}
\includegraphics[scale=0.33,angle=-90.0]{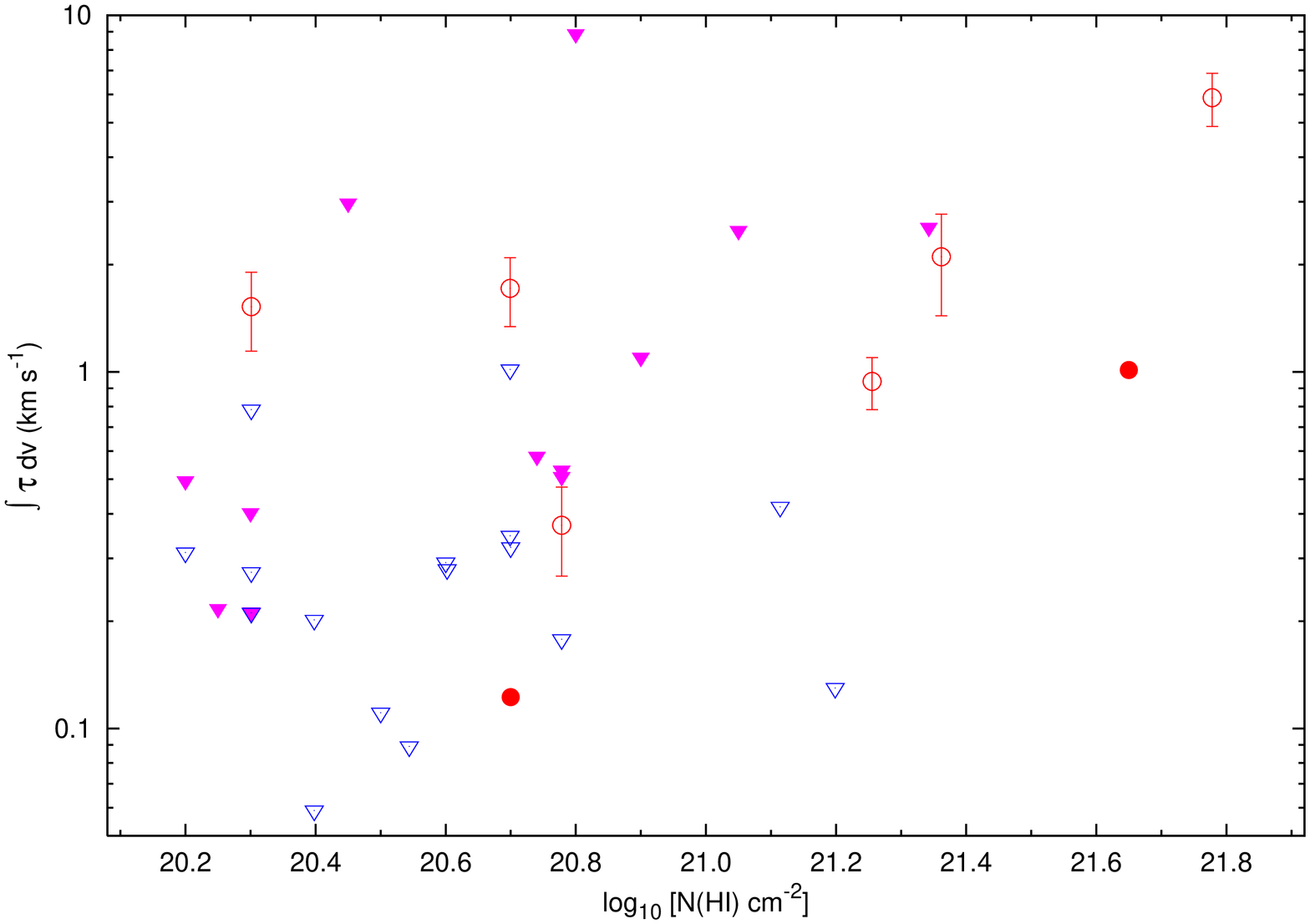}
\caption{\label{fig:fig4} Equivalent velocity width ($\int\tau{dv}$) measurements or upper limits, corrected for $f_c$, as a function of $z$ and N(H~{\sc i}). Legends are same as in Figure~\ref{fig:fig3} (triangles are inverted to indicate upper limits). For unknown $f_c$, the median value of $0.65$ is adopted. See \S\ref{sec:dis} for details.}
\end{center}
\end{figure*}

A comparison of our $T_{\rm s}$ limits with other $T_{\rm s}$ measurements and 
limits are shown in Figure~\ref{fig:fig3}. The $T_{\rm s}$ values with 
redshifts and H~{\sc i} column densities are shown in the left and the right 
panel respectively. The two filled circles, showing the limits from the 
present work, near the top right corner of the left panel are evidently very 
strong constraints on $T_{\rm s}$ at the high $z$ end. The dotted lines in 
both the panels of Figure~\ref{fig:fig3} approximately mark the lower envelope 
of the measured $T_{\rm s}$ values and limits; this suggests that the minimum 
$T_{\rm s}$ increases with both $z$ and N(H~{\sc i}). However, this conclusion 
comes dominantly from H~{\sc i} non-detections, and only few measurements. 
More such deep observations to constrain $T_{\rm s}$ tightly for more DLAs and 
to improve the statistical significance are certainly required to draw any 
definitive conclusions.

Finally, we note that in near future the new and upcoming low frequency 
telescopes will mediate a significant progress in this field. In particular, 
in about an year, the ongoing upgradation of the GMRT will offer a seamless 
frequency coverage of $\sim 50 - 1500$ MHz, and a slightly better sensitivity 
at $\lesssim 1000$ MHz due to improved low noise amplifiers (LNAs). These will 
make a larger redshift window available for H~{\sc i} 21 cm observations of 
DLAs, and will expand the horizon of DLA studies.

\section{Discussions}
\label{sec:dis}

\subsection{Redshift evolution or local origin?}

Redshift evolution of physical properties of the gas in DLAs has been a topic 
of much discussion. If real, the observed trend of higher average $T_{\rm s}$ 
at high $z$ will imply a lower fraction of cold gas at high redshift 
\citep[e.g.,][]{kan09b}, and may have important implications regarding cosmic 
evolution of properties/morphologies of DLA host galaxies. Thus, the issue is 
also related to the hierarchical galaxy formation models. Alternatively, 
\citet{cur05} attributed it to a lower covering factor at high $z$. On the 
other hand, based on C~{\sc ii}$^*$ observations, \citet{wol03} argued for the 
presence of a large fraction of cold gas in high $z$ DLAs, and suggested that 
the radio and optical lines of sight are completely different. However, 
\citet{kan09a, kan09b} argue that the observed trend is most likely {\it not} 
caused by the above two reasons, and is due to a real evolution of the cold 
gas fraction. 

One crucial argument in this context comes from the observed anti-correlation 
between $T_{\rm s}$ and metallicity \citep[e.g.,][]{kan09b,cur10,eli12}. 
\citet{sri12}, however, find no such anti-correlation, either due to possible 
redshift evolution or due to a small range of metallicity in their sample. 
\citet{kan09b} noted that a mass-metallicity relation in DLAs \citep[see, 
e.g.,][for details]{nee13} can naturally result in a larger cold gas fraction 
due to efficient cooling via metal line transitions for high metallicity 
systems, and leads to the observed anti-correlation. While this reinforces the 
evidence for an evolution of the cold gas fraction, \citet{kan09b} can not 
completely rule out the possibility of a local origin of the relation due to 
significant metallicity gradients -- a ``line-of-sight effect'' of higher 
metallicity and lower temperature for sightlines passing through the central 
regions and vice versa.

In this local origin scenario (i.e. if DLA hosts are large galaxies with 
internal metallicity and/or temperature gradients), one would expect to see an 
anti-correlation between the H~{\sc i} column density and the spin temperature 
as well. For example, in diffuse H~{\sc i} of the Milky Way, $T_{\rm s}$ is 
higher for lower N(H~{\sc i}) sightlines. This has been reported recently by 
\citet{kan11} from Galactic H~{\sc i} observations along $35$ lines of sight. 
A similar anti-correlation is also evident for a larger sample of $79$ lines 
of sight from the millennium Arecibo 21~cm absorption-line survey 
\citep{hei03}. We do not see such anti-correlation for DLAs. Rather, we see a 
weak correlation between N(H~{\sc i}) and $T_{\rm s}$ (see 
Figure~\ref{fig:fig3} right panel), which, we argue below, is probably a 
``secondary correlation''. The number of measurements, particularly at the 
high N(H~{\sc i}) end, in the current sample is too small to draw any stronger 
conclusion at this point. However, the lack of any N(H~{\sc i}) -- $T_{\rm s}$ 
{\it anti-correlation} indicates that local origin is an incorrect explanation 
(for the reported anti-correlation between $T_{\rm s}$ and metallicity).

\subsection{Primary vs. secondary correlations}

While discussing N(H~{\sc i}) -- $T_{\rm s}$ correlation, it is important to 
remember that $T_{\rm s}$ is a derived quantity. In Figure~\ref{fig:fig4} we 
instead show the measurements or upper limits of directly observed quantities, 
equivalent velocity width ($\int \tau {dv}$), with $z$ and N(H~{\sc i}). There 
is a weak indication of lower equivalent width at higher redshift, but no 
significant correlation with N(H~{\sc i}) is evident in this plot. So, the 
observed N(H~{\sc i}) -- $T_{\rm s}$ ``correlation'' is consistent with random 
equivalent velocity widths, and is most likely a secondary correlation that 
arises because $T_{\rm s}$ is derived from observables $\int \tau {dv}$ and 
{\it also} N(H~{\sc i}) itself. Note that for a sample with a similar redshift 
range, \citet{sri12} also found no $\int \tau {dv}$ -- N(H~{\sc i}) 
correlation, while \citet{cur10} reported a correlation for a sample with a 
larger redshift range. 

Is it possible that the redshift evolution of $T_{\rm s}$ is also a secondary 
correlation due to some selection effects? We have noted from 
Figure~\ref{fig:fig2}, that the current sample consists of both low and high 
N(H~{\sc i}) DLAs at high $z$, but selectively only low N(H~{\sc i}) DLAs at 
low $z$. The SDSS full sample does not have this selection effect. If 
$T_{\rm s}$ is approximately constant over the redshift range, one would then, 
for this sample, expect a lower value of equivalent width at low $z$, but a 
higher mean (and larger spread) at high $z$. In the data, we see an opposite 
trend -- a weak indication of an anti-correlation between redshift and 
equivalent width. So, it seems that the redshift evolution of the spin 
temperature (and/or of the equivalent width) is real, and not merely a 
selection effect. We note that the interdependence N(H~{\sc i}) $\propto$ 
$T_{\rm s} \times \int \tau {dv}$ makes it hard to decide, from the very 
limited sample, whether one of these correlations (spin temperature and 
equivalent width with redshift) is ``primary'', and the other one arises only 
as a secondary effect, or if both have real physical significance. There is 
some discussion in this regard by \citet{kan09b} based on the argument that a 
primary correlation is expected to have higher significance. More observations 
and careful comparison with theoretical models (e.g. with predicted spin 
temperature {\it and} velocity spread from the hierarchical galaxy formation 
models) are certainly required to address these issues.
 
\section{Conclusions}
\label{sec:con}

We have presented results from low frequency radio observations of two DLAs at 
$z > 3$. Based on non-detections of redshifted 21 cm absorption, the spin 
temperature $3\sigma$ limits are $> 2000$ K for both the systems. These are 
very strong limits for DLAs at such high redshift. A compilation from the 
literature of $T_{\rm s}$ measurements and limits for DLAs at $z > 2$ shows a 
clear indication of higher average gas temperature at high redshift. From the 
absence of N(H~{\sc i}) -- $T_{\rm s}$ anti-correlation, we conclude that the 
reported anti-correlation between $T_{\rm s}$ and metallicity is unlikely to 
be a local, line of sight effect. We also argue that the observed trend of 
higher $T_{\rm s}$ at high $z$ is neither a selection effect nor a derived 
secondary correlation, but a true, physical evolution. This implies a smaller 
fraction of cold gas in DLAs at high redshift, likely because of lower 
metallicity at high redshift. Access to a larger redshift range in near future 
for such studies will definitely be very useful.

\section*{Acknowledgments}
We thank Aritra Basu, Rahul Basu, Prasun Dutta, and the staff of the GMRT who 
have made these observations possible. GMRT is run by the National Centre for 
Radio Astrophysics of the Tata Institute of Fundamental Research. We are 
grateful to Jayaram N. Chengalur, Nissim Kanekar and Divya Oberoi for useful 
discussions. We also thank the editor and the anonymous reviewer for prompting 
us to improve this paper. NR acknowledges support from the Alexander von 
Humboldt Foundation and the Jansky Fellowship Program of the National Radio 
Astronomy Observatory (NRAO). The NRAO is a facility of the National Science 
Foundation operated under cooperative agreement by Associated Universities, Inc.

\bsp
\label{lastpage}
\end{document}